# Microresonator Soliton Frequency Combs in the Zero-Dispersion Regime

*Shuangyou Zhang,[1] Toby Bi,[1,2] and Pascal Del'Haye[1,2*]*

[1]*Max Planck Institute for the Science of Light, 91058 Erlangen, Germany*
[2]*Department of Physics, Friedrich-Alexander-Universität Erlangen-Nürnberg, 91058 Erlangen, Germany*
*[*]pascal.delhaye@mpl.mpg.de*

**Abstract:** Chip-scale optical frequency combs have attracted significant research interest and can be used in applications ranging from precision spectroscopy to telecom channel generators and lidar systems. In the time domain, microresonator based frequency combs correspond to self-stabilized soliton pulses. In two distinct regimes, microresonators have shown to emit either bright solitons in the anomalous dispersion regime or dark solitons (a short time of darkness in a bright background signal) in the normal dispersion regime. Here, we investigate the dynamics of continuous-wave-laser-driven soliton generation in the zero-group-velocity-dispersion (GVD) regime, as well as the generation of solitons that are spectrally crossing different dispersion regimes. In the measurements, zero-dispersion solitons with multi-peak structures (soliton molecules) are observed with distinct and predictable spectral envelopes that are a result of fifth-order dispersion of the resonators. Numerical simulations and the analysis of bifurcation structures agree well with the observed soliton states. This is the first observation of soliton generation that is governed by fifth-order dispersion, which can have applications in ultrafast optics, telecom systems and optical spectroscopy.

## 1. Introduction

Optical frequency combs based on monolithic high-Q microresonators have been intensively studied over the past 15 years [1–3]. In particular, their chip-scale footprint and low power consumption is of interest for out-of-the-lab applications. The discovery of dissipative Kerr solitons in microresonators has been demonstrated to enable low-noise, coherent and broadband frequency combs due to the balance between the Kerr nonlinearity, group velocity dispersion (GVD), cavity losses and cavity gain [4]. Rich nonlinear physics in microresonators have been revealed in the past decade, including breather solitons [5,6], soliton crystals [7,8], Stokes solitons [9], Pockels solitons [10], laser cavity solitons [11], dark solitons [12,13], and dark-bright soliton pairs [14]. In many of the aforementioned soliton types, the GVD of microresonators plays a critical role in microcomb formation. Soliton in microresonators have been separated into two distinct cases, bright solitons (intensity peaks on a low-level background) [4] or dark solitons (intensity dips embedded in a high-intensity background) [12], depending on the second order resonator dispersion at the pump wavelength. Bright soliton generation in microresonators requires anomalous GVD at the pump wavelength while dark solitons can be observed when pumping in the normal dispersion regime. Bright solitons can be described to originate from modulation instability [4], in contrast to dark solitons, which arise through the interlocking of switching waves connecting the homogenous steady-states of the bistable cavity system [15]. Recent research theoretically predicted the coexistence of bright and dark solitons in the regimes of normal [16], zero [17–19], and anomalous GVD [20], when taking account of higher-order dispersion (third- and fourth-order).

Among these studies, soliton generation in the zero GVD regime is of particular interest. Especially in conventional mode-locked lasers, dispersion compensation with prism pairs and chirped mirrors has been the key for the generation of femtosecond and attosecond pulses. Working in this regime, comb lines from different spectral sides relative to the pump laser experience opposite dispersion, normal dispersion on one side and anomalous dispersion on the other side. As a result, the solitons exhibit asymmetrical behavior both in frequency and time domain. Moreover, working at zero GVD allows to investigate the higher-order dispersion, which plays a dominant role in the soliton formation dynamics. The bright and dark soliton formation in this regime can also be described by the interlocking of switching waves [15]. Additionally, when pumping close to the zero-dispersion crossing, a dispersive wave (DW) is expected to be generated close to the pump, and hence, soliton formation will be strongly affected by the soliton recoil effect [17]. Zero or small GVD is the key to obtain spectrally broadband frequency combs, in particular with rapidly growing ways to control microresonator dispersion in waveguide structures. Just recently, Li *et al.* reported on the first experimental observations of bright soliton generation in a close-to-zero, weakly normal GVD fiber loop resonator [21]. Almost Anderson *et al.* demonstrate soliton formation pumping in near zero GVD (weakly normal dispersion) regime in chip-based $Si_3N_4$ microresonators, resulting in a near octave-spanning spectrum [22]. The zero-GVD solitons observed in these studies are enabled by third-order dispersion (TOD) and both studies use synchronously pulsed-driving of the cavities rather than a continuous wave (CW) laser. This synchronously pulsed-driving can be seen as quasi-cw, since the generated solitons are much narrower than the driving pulses. Especially in microresonator systems, pulsed pumping can strongly influence the soliton formation process and enables e.g. thresholdless comb generation [23].

In this paper, we demonstrate and explore the soliton dynamics in a microresonator pumped by a CW source in different dispersion regimes ranging from anomalous, crossing zero, to normal GVD. Our studies reveal that the close-to-zero-dispersion solitons in our system are enabled by the fifth-order dispersion (5th-OD). This is in contrast to the previous studies [16,17,21,22], in which zero-dispersion solitons are induced by the TOD. Different zero-dispersion soliton states can be accessed with a predictable spectral profile by adiabatically tuning the laser frequency and pumping different optical modes. To the best of our knowledge, this is the first observation of zero-GVD solitons in a microresonator driven by a CW laser, and more importantly, the first report of 5th-OD induced soliton.

## 2. Experimental setup

Figure 1(a) shows a schematic of the experimental setup. The pump laser at 1.3 μm wavelength is used to generate bright soliton structures to study the soliton dynamics in different GVD regimes. Another laser at 1.5 μm wavelength is used as an auxiliary laser to passively stabilize the circulating optical power within the microresonator to assist soliton generation for the 1.3-μm pump laser [24]. A 250-μm-diameter fused silica microtoroid with an FSR of 257 GHz is used in the experiments and shown in the inset of Fig. 1(a). This particular microresonator is fabricated from a silicon wafer with a 6-μm layer of thermally grown silicon dioxide ($SiO_2$) [25]. The two lasers are combined with a wavelength division multiplexer (WDM) and evanescently coupled into the microresonator via a tapered optical fiber. Two fiber polarization controllers (PCs) are used to match the polarization of the two lasers to the respective cavity mode polarizations. At the resonator output, the light of the two wavelengths is separated by another WDM. Part of the 1.3-μm soliton comb light is monitored with an optical spectrum analyzer (OSA) and a photodiode (PD). An autocorrelator (AC) based on second-harmonic generation is used to measure the autocorrelation traces of the 1.3-μm comb light. In addition, a fiber Bragg grating (FBG) filter is used to suppress the pump power before the AC. In the experiments, the auxiliary laser is used to pump a mode at 1556 nm for the thermal stabilization of the resonator.

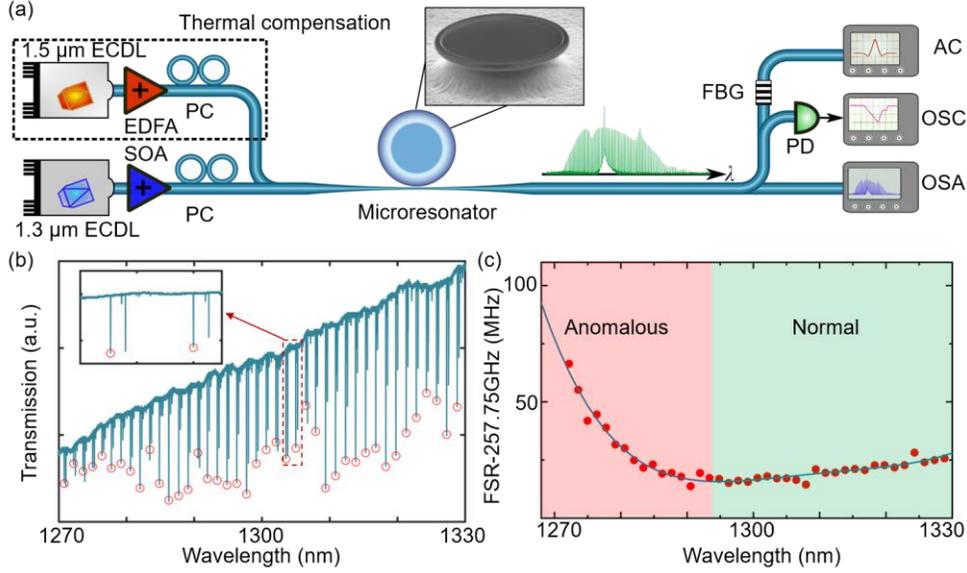

Fig. 1. Experimental demonstration of soliton dynamics crossing different dispersion regimes. (a) Experimental setup. The 1.3-µm pump laser is used to generate bright solitons by pumping a fused silica microtoroid whose GVD evolves from anomalous to normal within the wavelength range from 1270 to 1330 nm. The 1.5-µm auxiliary laser thermally stabilizes the resonator during the 1.3-µm soliton generation. ECDL: external cavity diode laser; EDFA: erbium-doped fiber amplifier; SOA: semiconductor optical amplifier; PC: polarization controller; PD: photodetector; AC: autocorrelator; OSA: optical spectrum analyzer; OSC: oscilloscope; FBG: fiber Bragg grating. Inset: scanning electron microscope image of the microtoroid used in the experiments. (b) Measured mode spectrum of the microtoroid. Inset shows the zoomed-in spectrum around 1304 nm. (c) Dispersion of the mode family marked with red circles in (b). Red circles are experimental data while the solid blue trace is a 4$^{th}$-order polynomial fit, considering up to the fifth-order dispersion.

The resonance frequencies of a mode family in a whispering-gallery mode resonator made of a dispersive medium can be described as

$$\omega_\mu = \omega_0 + D_1\mu + \frac{D_2}{2!}\mu^2 + \frac{D_3}{3!}\mu^3 + \frac{D_4}{4!}\mu^4 + \frac{D_5}{5!}\mu^5, \quad (1)$$

where $\mu$ is the mode number offset from the pump mode at $\mu = 0$ and $\omega_\mu$ are the resonance frequencies. $D_1/2\pi$ is the FSR of the resonator at the pump mode ($\mu = 0$), and $D_2$, $D_3$, $D_4$, and $D_5$ are coefficients of second-, third-, fourth-, and fifth-order dispersion, respectively. Figure 1(b) shows the measured mode spectrum of the microtoroid used here in the region between 1270 nm and 1330 nm. There are three different mode families observed in the transmission trace (See Fig. S1 in Supplement 1). The mode family (red circles) with an intrinsic optical quality factor of 30 million is selected. The GVD of this mode family changes from anomalous to normal. Figure 1(c) shows the corresponding FSR evolution of the selected soliton mode family. In this measurement, we confirm that the variation of the FSR with respect to the wavelength changes its sign. The GVD of the optical modes shorter than 1294 nm is anomalous (FSR increases with optical frequency), while it becomes normal for wavelengths larger than 1294 nm (FSR decreases with optical frequency). Even though some experimental data points in Fig. 1(c) deviate from the fitted curve, we believe these modes are not related to avoided mode crossings (See Supplement 1 for details). In the experiments, the 1.3-µm pump laser is tuned from 1270 nm to 1304 nm, correspondingly, the GVD of the soliton modes changes from anomalous to normal. Within this range, each longitudinal resonator mode (from the same mode family that is marked with red circles in Fig. 1(b)) supports the formation of a bright soliton, but shows significantly different dynamics due to the variation of the GVD.

## 3.  Bright soliton generation with asymmetric dispersion

First, we start by pumping a mode at 1283.4 nm in the anomalous GVD regime with a pump power of 55 mW. By optimizing the power and laser-cavity detuning of the 1.5-µm auxiliary laser, 1.3-µm soliton states can be deterministically accessed by manually tuning the 1.3-µm pump laser across its resonance from the blue-detuned side to the red-detuned side [24]. Fig. 2(a) shows the measured integrated dispersion profile (blue circles) $D_{int}(\mu)= \omega_\mu - \omega_0 - \mu D_1$ of this pump mode (1283.4 nm), together with a 5$^{th}$-order polynomial fit (blue curve). The dispersion trace shows anomalous GVD ($D_2/2\pi$ = 2.1 MHz, $D_3/2\pi$ = 424 kHz, $D_4/2\pi$ = 48.3 kHz, and $D_5/2\pi$ = 2.8 kHz) but also an asymmetric profile on either side of the pump mode. On the left side of the pump wavelength, the dispersion is much stronger than that on the right side, which can be seen from the sharper slope on the left side. As a result, the non-coherent, chaotic spectrum shown in Fig. 2(b) shows a strong asymmetry with a relatively narrow spectrum on the lower wavelength side and broad spectrum on the higher wavelength side. Subsequently tuning the 1.3-µm pump laser to the red-detuned side of its resonance, a bright single soliton can be generated as shown in Fig. 2(c), with a sech$^2$ envelope and a DW at 1388 nm.

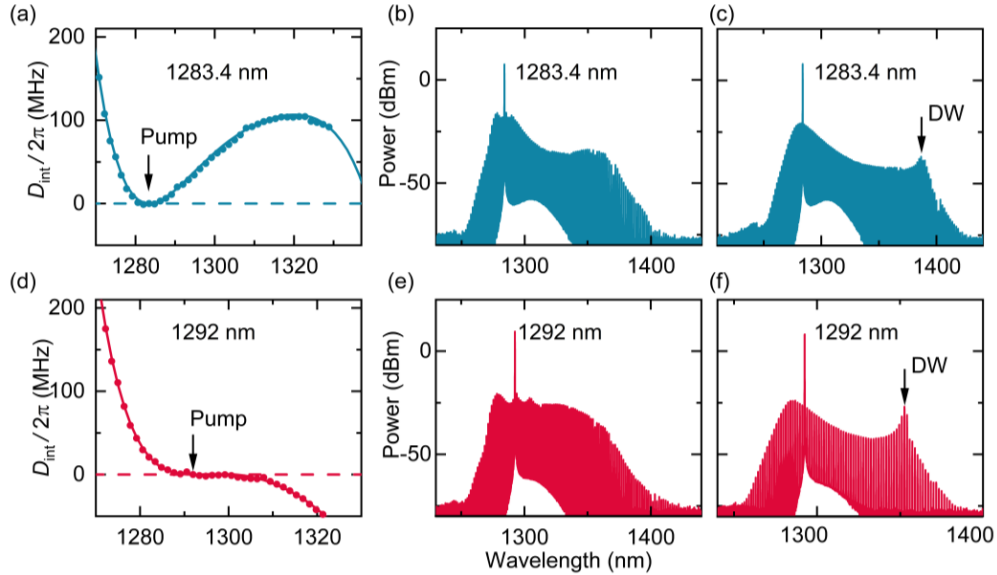

Fig. 2. Bright soliton generation with asymmetric dispersion. (a) Measured integrated dispersion profile (blue circles) at pump wavelength of 1283.4 nm together with a polynomial fit (solid trace). Optical spectrum when the microcomb is in a chaotic (b) and single-soliton state (c). (d) Measured integrated dispersion profile (red circles) at pump wavelength of 1292 nm together with a polynomial fit (solid trace). Optical spectrum when the microcomb is in a chaotic (e) and single-soliton state (f).

In the next step, we further increase the pump wavelength to approach the zero-GVD regime. Figure 2(d) shows the integrated dispersion measurement (red circles) at the pump wavelength of 1292 nm together with the fitted curve (red). Here, the local dispersion around the pump laser has opposite signs, transitioning from positive values for $D_{int}$ to negative values for increasing wavelengths. The calculated $D_2/2\pi$, $D_3/2\pi$, $D_4/2\pi$, and $D_5/2\pi$ at this pump wavelength are 307.8 kHz, 184.2 kHz, 31.6 kHz, and 2.8 kHz, respectively. Figure 2(e) and 2(f) show the comb spectra with a chaotic state (Fig. 2(e)) and a single-soliton state (Fig. 2(f)). The single soliton spectrum in Fig. 2(f) has one DW around 1353 nm, which is not phase-matched according to the dispersion profile in Fig. 2(d). After the soliton formation, the mismatch between the repetition rate and the local FSR creates phase-matching for the DW in Fig. 2(f). This is attributed to the influence of the dependence of the repetition rate on the soliton recoil, Raman shift, and the laser-cavity detuning [26,27]. See section 2 in Supplement 1 for details. As a result of the soliton recoil effect [28], the appearance of the DW on the red side of the spectrum causes a blue shift of the central wavelength of the spectral envelope with respect to the pump wavelength. Comparing both spectra in Fig. 2(c) and (e), indeed, soliton formation is strongly affected by the soliton recoil effect when pumping close to the zero-dispersion regime.

## 4. Soliton generation at zero GVD

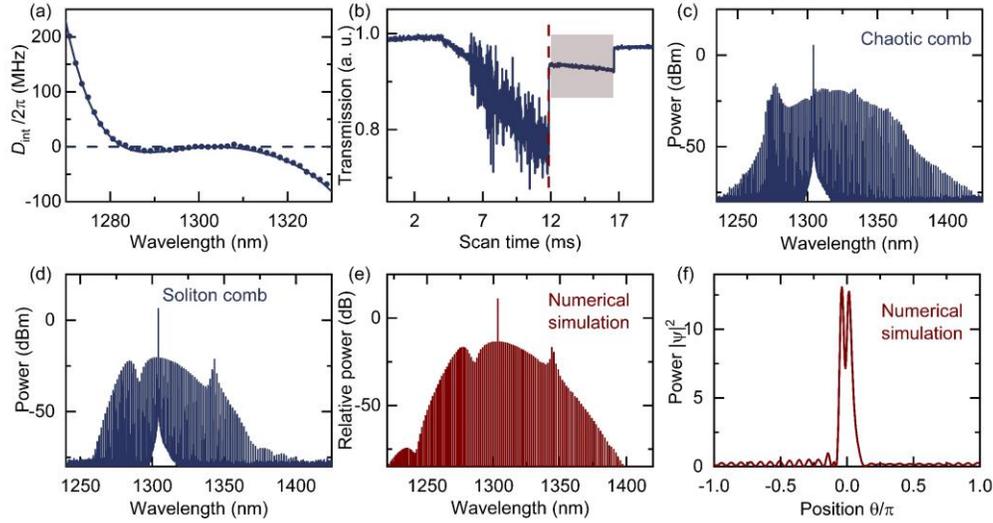

Fig. 3. Measurement of a zero-GVD soliton in a microresonator driven by a continuous wave laser. (a) Integrated dispersion measured at 1303.6 nm. (b) Measured transmission trace when scanning the 1.3-μm pump laser frequency from the blue-detuned to the red-detuned side of the resonance. (c) Measured spectrum when the microcomb is in a chaotic state. (d) Measured spectrum of a soliton state. (e) Numerically simulated optical spectrum and the corresponding intracavity temporal waveform (f). $|\psi|^2$ is the dimensionless intracavity power.

To explore the soliton dynamic in the close-to-zero GVD regime, we tune the pump wavelength to slightly longer wavelengths. Figure 3(a) shows the integrated dispersion measurement at the new pump wavelength of 1303.6 nm together with a fifth order polynomial fit of the data. The calculated $D_2/2\pi$, $D_3/2\pi$, $D_4/2\pi$, and $D_5/2\pi$ at this pump wavelength are −390.4 kHz, 19.9 kHz, 9.5 kHz, and 2.8 kHz, respectively. When the pump laser approaches the zero-GVD regime, the dispersion close to the pump mode becomes flatter compared to Fig. 2(a) and 2(d). Figure 3(b) shows a transmission trace when slowly scanning the 1.3-μm pump laser across the resonance from the blue- to the red-detuned side while simultaneously coupling the 1.5-μm auxiliary laser into its resonance for thermal compensation. On the blue-detuned side of the soliton resonance (left side of the dashed line), the comb is in a non-coherent chaotic state and the corresponding spectrum is shown in Fig. 3(c). Note that, even though under normal dispersion, the field experiences modulation instability gain, which is enabled by the fourth-order dispersion (4th-OD) of our system. The 4th-OD has an opposite sign compared to the second-order dispersion, making the phase-matching condition possible for four-wave mixing [29–32]. This enables the generation of bright solitons under normal dispersion with an adiabatic laser detuning scan. As expected from Fig. 3(a), the spectrum becomes more symmetric with respect to the pump laser, in comparison to the spectra in Fig. 2(b, e). We can still observe a distinct DW at 1277 nm.

When tuning the pump laser frequency into the soliton step (shaded area in Fig. 3(b)), a bright soliton is formed with a significantly smoother spectral envelope compared to the chaotic spectrum in Fig. 3(c). Figure 3(d) shows the optical spectrum of this bright zero-GVD soliton. Interestingly, the soliton spectrum has a distinct envelope with a spectral dip on the blue-detuned side of the pump and a DW on the red-detuned side. The distinct envelope is different from these of conventional bright (sech$^2$ envelope) and dark solitons. To investigate the intracavity dynamics of the zero-GVD soliton, a number of numerical simulations are performed based on a generalized mean-field Lugiato-Lefever equation (LLE) [33–35], taking into account up to 5th-OD (See Supplement 1 for details about the numerical model). Figure 3(e) and (f) show the numerically simulated optical spectrum and the corresponding intracavity temporal waveform, respectively. The simulated spectra are in excellent agreement with the measurements. The intracavity temporal waveform in Fig. 3(f) reveals a two-peak bright soliton with a DW-induced oscillation in the pedestals.

Theoretical studies [16,17] predicted that TOD can enable the coexistence of dark and bright solitons under normal dispersion. Recent work has demonstrated that zero-dispersion solitons with multiple peaks are induced by TOD [21,22]. In contrast, the TOD in our system is too weak to support bright solitons or multi-peak bright solitons. To verify the dominant effect of 5th-OD, we theoretically investigate the bifurcation structure and stability of the soliton states in our system. By solving the steady-state solutions of the LLE using a Newton-Raphson continuation algorithm [34], Figure 4 shows three collapsed snaking bifurcation structures [16,21] under different dispersion configurations. Figure 4(a) presents the snaking bifurcation structure of the energy of bright solitons as a function of laser detuning with dispersion parameters of $D_2/2\pi = -390.4$ kHz, $D_3/2\pi = 19.9$ kHz, $D_4/2\pi = 9.5$ kHz, and $D_5/2\pi = 2.8$ kHz, where several different stable soliton solutions (solid trace) coexist in the system and where each solid line at different energy levels corresponds to different bright soliton states with different peak numbers in the temporal profile, as shown in the first column of Fig. 4d. The detuning range of two-, three-, and four-peak solitons is relatively large. However, we also observe single-peak solitons in the simulation, suggesting that the two-peak solitons in Fig. 3 can be understood as two solitons that are bound together, which can be referred to as a soliton molecule [36,37]. In this case, the binding mechanism is determined by the 5th-OD of the resonator. On the other hand, the measured two-peak soliton can be possibly explained as interlocked switching waves, where the up-switching wave has two peaks on the high-intensity solution induced by 5th-OD. Note that, with the particular pump wavelength used in this measurement, accessing a single-peak soliton is challenging due to the relatively narrow detuning range of its existence, as shown in Fig. 4(a).

In comparison, Figures 4(b) and 4(c) show the bifurcation diagram with dispersion parameters of $D_2/2\pi = -390.4$ kHz, $D_3/2\pi = 19.9$ kHz, $D_4/2\pi = 9.5$ kHz, and $D_5/2\pi = 0$ kHz for (b), and $D_2/2\pi = -390.4$ kHz, $D_3/2\pi = 25.3$ kHz, $D_4/2\pi = 0$ kHz, and $D_5/2\pi = 0$ kHz for (c), respectively. Figure 4(b) shows the bifurcation diagram in the presence of second-, third- and fourth-order dispersion. Without 5th-OD, there is only one bright soliton solution in the system, with a temporal profile shown in the middle column of Fig. 4(d). As discussed above, the 4th-OD in our system enables the generation of bright solitons by inducing phase-matching for four-wave mixing in the zero GVD region. However, it is not responsible for the observed zero-GVD solitons. Figure 4(c) shows the bifurcation diagram under the impact of only $D_2$ and $D_3$. Please note that under normal dispersion, the existence range of bright solitons increases with the value of $D_3$ [16]. In our system, the experimental value $D_3/2\pi$ is weak (19.9 kHz, corresponding to −0.17 in dimensionless units) and hinders the numerical simulation from finding stable solutions. To illustrate the effect of TOD on the bifurcation structure, we increase $D_3/2\pi$ to 25.3 kHz, corresponding to a dimensionless TOD of −0.22. As shown in Fig. 4(c), there are several stable solutions of bright solitons with different peak numbers coexisting in the system, whose temporal profiles are plotted in the third column of Fig. 4(d). However, there are no stable single-, two- and three-peak solitons. The first stable bright soliton in the system is a four-peak soliton with an existence range of only 25% of a resonance linewidth. With the bifurcation results in Fig. 4, we conclude that the existence range of multi-peak solitons is greatly extended by the 5th-OD, and therefore, the zero-dispersion solitons observed here are enabled by 5th-OD. See Supplement 1 for further details on the impact of TOD, 4th-OD, and 5th-OD on the soliton generation.

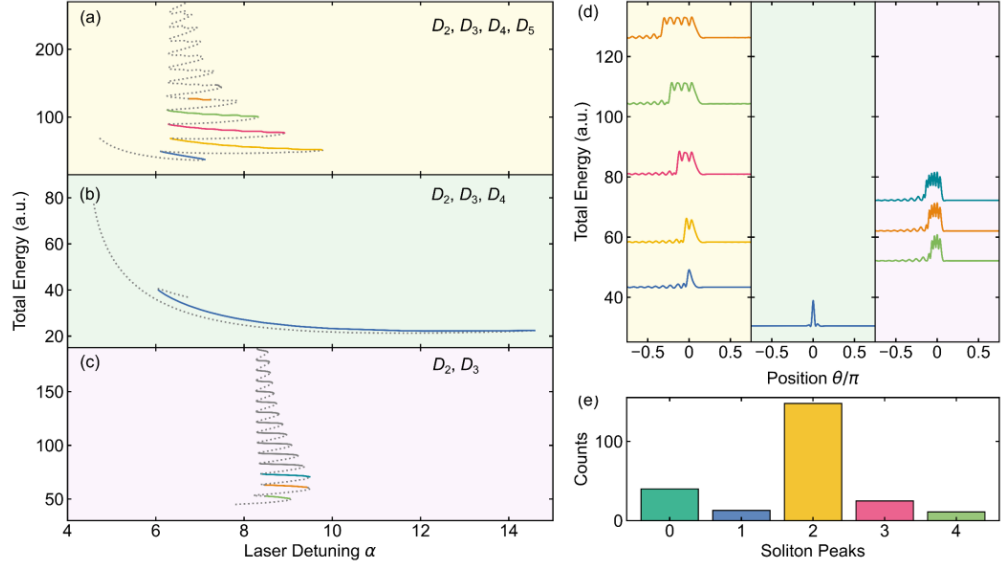

Fig. 4. Numerically simulated snaking bifurcation structure under different dispersion parameters as a function of laser detuning $\alpha$ (normalized to half of a cavity linewidth). (a), Simulation with dispersion parameters of $D_2/2\pi$, $D_3/2\pi$, $D_4/2\pi$, and $D_5/2\pi$ of -390.4, 19.9, 9.5, and 2.8 kHz. (b), Simulation with dispersion parameters of $D_2/2\pi$, $D_3/2\pi$, and $D_4/2\pi$, of -390.4, 19.9, and 9.5 kHz. (c), Simulation with dispersion parameters of $D_2/2\pi$, and $D_3/2\pi$ of -390.4 and 25.3 kHz. Note that the $D_3$ used for this simulation is higher than the experimentally measured value due to the difficulty in obtaining stable bright-soliton structures at lower TOD. (d), Corresponding stable bright-soliton temporal profiles of (a), (b), and (c) offset by the total energy in the cavity. (e), Histogram of peak numbers of numerically simulated solitons when sweeping the laser detuning using the dispersion parameters from panel (a).

Figure 4(a) suggests that multi-peak soliton states with different temporal and spectral profiles can coexist under the same system parameters with the influence of the 5th-OD (See first column in Fig. 4(d) and Supplement 1 for details). However, apart from the two-peak soliton in Fig. 3, other multi-peak soliton states are not observed in our experiments when adiabatically scanning the laser detuning. The reason is that the existence range of the two-peak solitons is larger than that of other multi-peak soliton states. When slowly changing the laser frequency from the blue- to the red-detuned side of the resonance, the intracavity field experiences intense and random chaotic states, and the spontaneous formation of bright soliton structures tends to be the state with the highest probability (two-peak soliton in our case). To further verify the two-peak soliton as the highest probability state, Figure 4(e) shows the histogram of peak numbers of the generated soliton states when repeatedly running a split-step LLE simulation by scanning the laser frequency across the cavity resonance using the same parameters as used in Fig. 3. Note that, once generated, the zero-GVD solitons demonstrated in this work are stable and sustain their shape within a large cavity-pump detuning range. See Supplement 1 for details.

The results presented above demonstrate that zero-GVD solitons at a specific pump mode can be generated with a predictable profile and sustain its shape regardless of cavity-pump detuning variation. Here, we highlight that several types of zero-GVD solitons with different temporal and spectral profiles can be accessed by pumping different optical modes. Figure 5(a) shows two measurements of the transmission signal when scanning the 1.3-μm pump laser across a resonance at 1299.2 nm, with dispersion parameters of $D_2/2\pi = -275.8$ kHz, $D_3/2\pi = 60.7$ kHz, $D_4/2\pi = 17.8$ kHz, and $D_5/2\pi = 2.8$ kHz. In the red-detuned regime, we observe two different soliton steps. Figure 5(b) shows an experimental spectrum of a bright soliton when the microcomb is in the lower soliton step (yellow area in Fig. 5(a)). It is interesting to note that, in contrast to the experimental spectra in Fig. 3, there is another spectral dip on the longer-wavelength side of the pump, in addition to the one on the shorter-wavelength side. Similar to Fig. 3, there is also a DW on the longer-wavelength side of the pump laser. Figure 5(c) shows the numerically simulated spectrum, which agrees well with the measured spectrum in Fig. 5(b). Figure 5(d) shows the simulated intracavity temporal waveform corresponding to the spectrum in Fig. 5(c), which suggests that the experimental result in Fig. 5(b) corresponds to a bright doublet soliton in the time domain. To further verify the temporal profile, we measure the autocorrelation traces by sending the light with the spectrum from Fig. 5(b) into an autocorrelator. An FBG filter is used to suppress the pump laser and a semiconductor optical amplifier amplifies the comb signal before sending the light into the autocorrelator. The filtered and amplified comb spectrum is presented in the Supplement 1. Figure 5(e) shows the normalized autocorrelation trace of the two-peak soliton. The two side-lobes are associated with the doublet structure shown in Fig. 5(d). Compared with Fig. 3(f), the temporal delay between the two peaks in Fig. 5(d) is larger, further suggesting that the two-peak soliton demonstrated here can be considered as soliton molecule, where two individual solitons are bound together. In fact, when pumping this mode, a single isolated bright soliton can also be accessed experimentally. The corresponding spectrum is shown in Fig. 5(f) when the microcomb is in the upper soliton step (blue area) in Fig. 5(a).

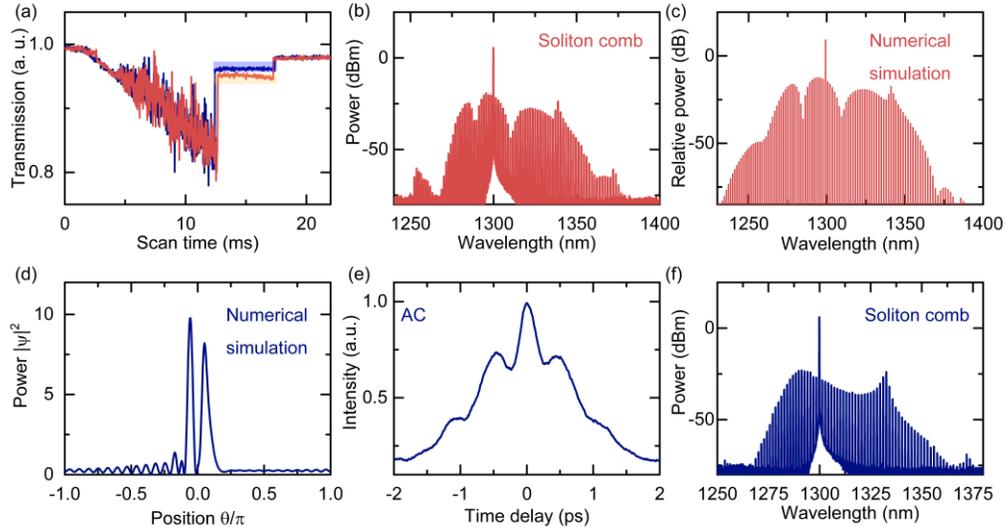

Fig. 5. Measurement of different zero-GVD solitons driven at different microresonator modes. (a) Experimental transmission traces when scanning the 1.3-μm pump laser frequency across its resonance from the blue- to the red-detuned side. (b) Measured spectrum of a zero-GVD soliton at 1299.2 nm. (c) Numerically simulated intracavity optical spectrum and the corresponding temporal waveform (d). (e) Measured autocorrelation trace (blue) of the zero-GVD soliton with the spectrum shown in (b). (f) Spectrum of a single isolated bright soliton in the zero-GVD regime.

## 5. Conclusions

In conclusion, we explored microresonator soliton dynamics in the zero-GVD regime and in the transition range between anomalous and normal dispersion, driven by a continuous wave laser. Numerical simulation and investigations of bifurcation structure reveal that the observed zero-GVD solitons are enabled by fifth-order dispersion. This is the first demonstration of soliton formation induced by fifth-order dispersion. Multi-peak soliton structures can be accessed with a stable temporal profile when pumping optical modes close-to-zero GVD. The shape and temporal delay of the zero-GVD solitons can be controlled by changing the pump modes. These results could be used for new types of chip-integrated pump-probe-experiments. In particular with the rapid advances in microfabrication techniques for waveguide dispersion control, this work can help to generate broadband optical frequency combs in dispersion compensated microresonators with applications in precision spectroscopy, optical telecommunication systems [37], and chip-integrated lidar systems.

**Funding.** European Union's H2020 ERC Starting Grant "CounterLight" 756966; H2020 Marie Sklodowska-Curie COFUND "Multiply" 713694; Marie Curie Innovative Training Network "Microcombs" 812818; and the Max Planck Society.

**Disclosures.** The authors declare no conflicts of interest.

**Data availability.** Data underlying the results presented in this paper are available from the authors upon reasonable request.

# Supplemental Information

**Dispersion of microresonators used in this study**

The microtoroids used in our study have a minor diameter of ~12-micrometer, which results in many different mode families coexisting in the resonator [1]. Depending on the diameter of the tapered fiber used for coupling and its relative position to the resonator, different mode families can be exited. In our experiments, the thickness of the tapered fiber at the coupling region and the coupling position are well selected, in order to archive a sufficient coupling efficiency for the zero-dispersion mode family, as well as to reduce the number of coupled mode families. The mode spectrum in Fig.1(b) in the main text shows 3 mode families observed in the experiments. Figure S1 shows the FSR evolution of the 3 mode families, in which the modes marked with blue circles are used for the zero-dispersion soliton generation (same modes in Fig. 1(c) in the main text). The modes marked with black triangles can only be measured up to 231 THz because their coupling becomes weaker as the wavelength decreases, preventing the observation of these modes in the optical spectrum. Fig. S1 does not any disturbances from other mode families onto the zero-dispersion soliton modes. In addition, the envelopes of all experimental spectra in this work do not show distortions induced by avoided mode crossings in the measurement range (1270 nm to 1330 nm). Therefore, we believe the formation of the zero-dispersion solitons does not depend on avoided mode crossings. This is also supported by the numerical simulations of soliton states without mode crossings, which show good agreement with the experimental results.

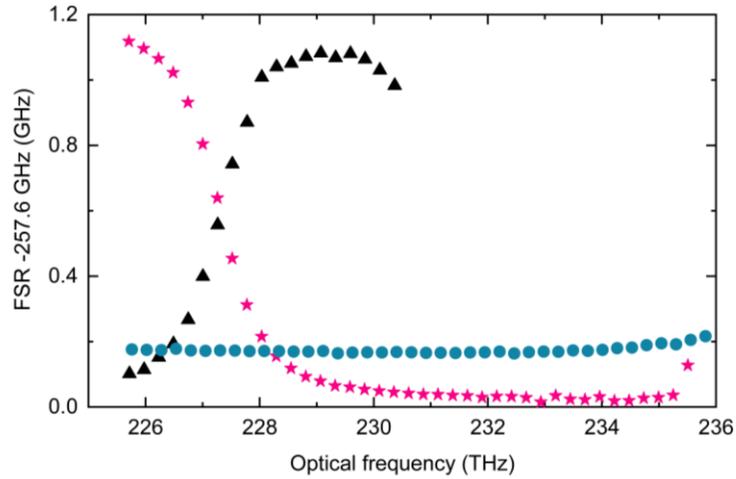

**Fig. S1** FSR evolution of 3 mode families measured in Fig. 1(b) in the main text. Modes marked with blue circles are the mode families used for the investigation of zero-dispersion solitons. The modes marked with black triangles can be measured up to 231 THz because the coupling becomes weaker as the mode wavelength gets smaller.

**Dispersive wave in the microcomb**

A dispersive wave (DW) in a microcomb is expected to occur at a mode number $\mu'$ at which the comb mode is exactly on resonance with the resonator mode. Thus $\mu'$ satisfies the condition

$$\omega_p + \omega_{rep}\mu' = \omega_0 + D_1\mu' + D_{int}(\mu'). \tag{S1}$$

Here, $\omega_p$ is the pump laser frequency and $\omega_{rep}$ is the repetition rate of the frequency comb. $\omega_0$ is the pump mode frequency, $D_1/2\pi$ is the FSR of the resonator at the pump mode, and $D_{int}(\mu')$ is the integrated dispersion relative to the pump mode. Eq. S1 can be rearranged to

$$D_{int}(\mu') = (\omega_{rep} - D_1)\mu' + \omega_p - \omega_0. \tag{S2}$$

Therefore, the phase matching condition for a DW is determined by ($\omega_{rep} - D_1$) and the pump laser detuning. The soliton repetition rate is determined by

$$\omega_{rep} = D_1 + D_2\frac{\Omega}{D_1} + \frac{D_3}{2!}\left(\frac{\Omega}{D_1}\right)^2 + \frac{D_4}{3!}\left(\frac{\Omega}{D_1}\right)^3 + \frac{D_5}{4!}\left(\frac{\Omega}{D_1}\right)^4, \tag{S3}$$

where $\Omega$ is the frequency shift between the pump and the soliton spectral maximum, which is a result of the soliton recoil from DWs and from the Raman-induced soliton self-frequency shift [2]. The total shift is given by $\Omega = \Omega_{Raman} + \Omega_{Recoil}$. The Raman-induced soliton self-frequency shift sets the soliton spectral center on the red side of the pump laser. From the soliton spectrum in Fig. 2(f), the soliton's spectral maximum is strongly blue shifted due to the recoil effect from the DW at 1353 nm. Therefore, the soliton recoil is the dominant effect in our measurements. Based on the measured dispersion parameters ($D_2/2\pi = 307.8$ kHz, $D_3/2\pi = 184.2$ kHz, $D_4/2\pi = 31.6$ kHz, and $D_5/2\pi = 2.8$ kHz) at this pump mode (1292 nm), the resulting $\omega_{rep}$ is larger than $D_1$. Consequently, the slope of the right-hand side of Eq. S2 is positive with respect to the mode number. Considering that the pump laser is in the red-detuned regime for soliton states, the right-hand side of Eq. S2 can be interpreted by the dashed line plotted in the figure below, creating a phase matching point (marked with an arrow) at a mode number of −40, which corresponds to the DW at 1353 nm.

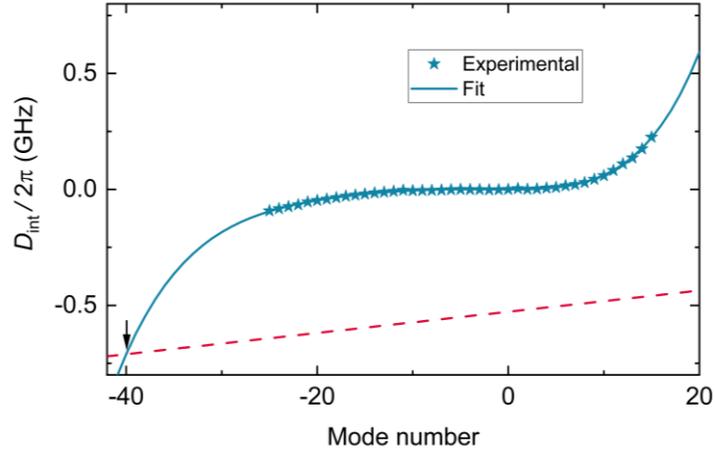

**Fig. S2** Measured integrated dispersion profile (blue stars) at a pump wavelength of 1292 nm together with a polynomial fit (solid trace). The crossing point (marked with an arrow) between the dashed line and $D_{int}$ shows the phase-matching point for the DW at 1353 nm in Fig. 2(f) in the main text.

**Numerical simulations of zero-dispersion solitons in a microresonator**

To simulate the temporal dynamics of the formation of zero-dispersion solitons in a microresonator, we solve a generalized Lugiato-Lefever equation (LLE) [3,4]:

$$\frac{\partial \psi(\theta,\tau)}{\partial \tau} = -(1+i\alpha)\psi + i|\psi|^2\psi - \sum_{n=2}^{N\geq 2}(-i)^{n+1}\frac{\beta_n}{n!}\frac{\partial^n \psi}{\partial \theta^n} + F, \tag{S4}$$

where $\tau$ is the slow time, normalized to twice the photon lifetime ($\tau_{ph}$) and $\theta$ is the azimuthal angle in a frame co-rotating at the group velocity. $\psi(\theta,\tau)$ is the intracavity field envelope. $\alpha$ is the frequency detuning of the pump laser (angular frequency $\omega_p$) with respect to the resonance frequency (angular frequency $\omega_0$) and normalized to half of the full-width at half-maximum (FWHM) of the resonance $\Delta\omega_0$,

$$\alpha = -\frac{\omega_p - \omega_0}{\Delta\omega_0 / 2}. \tag{S5}$$

The coefficient $\beta_n$ is the $n^{th}$-order normalized and dimensionless dispersion coefficient at the primary pump mode with $\beta_n = -2D_n/\Delta\omega_0$. The constant $F$ is the dimensionless external pump amplitude [4], normalized as

$$F = \sqrt{\frac{8g_0 \Delta\omega_{ext}}{\Delta\omega_0^3} \frac{P}{\hbar\omega_p}}, \tag{S6}$$

where $P$ is the input optical power in Watt. The constant $\Delta\omega_{ext}$ is the coupling linewidth. The nonlinear gain $g_0$ is given as

$$g_0 = \frac{n_2 c \hbar \omega_p^2}{n_0^2 V_0}, \tag{S7}$$

where $n_2$ and $n_0$ are the nonlinear (second-order) and linear indices of refraction, respectively. The constant $c$ is the vacuum speed of light. $V_0$ is the effective nonlinear mode volume. The LLE simulations are solved numerically using the split-step Fourier method and the Newton-Rapson method.

**Impact of third-order dispersion**

For the case of the pump laser at a wavelength of 1303.6 nm, the dispersion parameters are $D_2/2\pi$, $D_3/2\pi$, $D_4/2\pi$, and $D_5/2\pi$ are −390.4 kHz, 19.9 kHz, 9.5 kHz, and 2.8 kHz, respectively. Figure S3 shows a numerical simulation of the generation of a dark pulse when using the second- and third-order dispersion ($D_2/2\pi$ = -390.4 kHz, $D_3/2\pi$ = 19.9 kHz, $D_4/2\pi$ = 0 kHz, $D_5/2\pi$ = 0 kHz). In the simulation, the external pump amplitudes and laser detuning for the simulations are $|F|^2 = 12$ and $\alpha = 7.5$, respectively. Even though the initial condition corresponds to a bright pulse, after a duration of 300 photon lifetimes, the system eventually evolves into a stable dark pulse. Figure S3 (b) and (c) show the temporal profile and the corresponding spectrum.

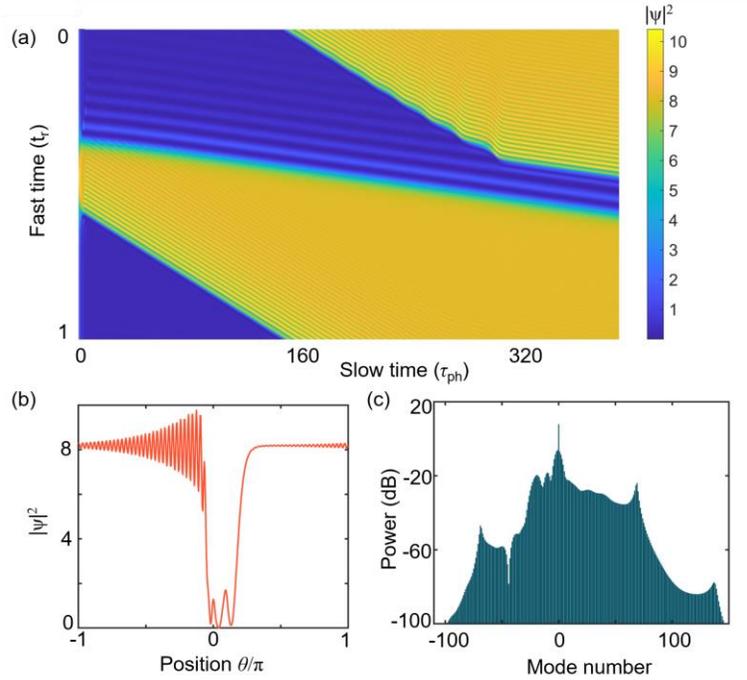

**Fig. S3** Numerical simulation of the generation of a dark pulse when the third-order dispersion is relatively weak ($D_2/2\pi$ = -390.4 kHz, $D_3/2\pi$ = 19.9 kHz, $D_4/2\pi$ = 0 kHz, $D_5/2\pi$ = 0 kHz). The external pump amplitudes and laser detuning for the simulations are $|F|^2 = 12$, $\alpha = 7.5$, respectively. (a) Evolution of an initially bright pulse to a dark pulse. (b) Temporal profile of the final stable dark pulse and its corresponding optical spectrum shown in panel (c).

**Impact of fourth-order dispersion**

The fourth-order dispersion of our resonators fundamentally alters the comb generation dynamics. The $D_4$ (9.5 kHz) has an opposite sign with respect to the second-order dispersion $D_2$ (−390.4 kHz), which makes the phase-matching condition for four-wave mixing and modulation instability possible in the normal dispersion regime [5]. Figure S4 below shows the evolution of the comb dynamics altered by $D_4$ with the dispersion parameters $D_2/2\pi$ = −390.4 kHz, $D_3/2\pi$ = 19.9 kHz, $D_4/2\pi$ = 9.5 kHz, $D_5/2\pi$ = 0 kHz, and pump power $|F|^2 = 12$. When the laser detuning is scanned from the blue-detuned to the red-detuned side of the resonance, the temporal intracavity field shown in Fig. S4 (a) generates a first pair of comb side bands at the phase-matched modes (± 28), then undergoes a Turing pattern, and finally forms a bright soliton. Figure S4(b) shows the spectral evolution of the intracavity field. Figure S4 (c) and (d) show the temporal profile (c) and optical spectrum (d) of the final stable bright soliton at the detuning marked with an arrow in Fig. 4(a). As a result, the impact of the fourth-order dispersion hinders the generation of dark solitons in the zero dispersion regime [6,7]. We must emphasize that the spectrum shown in Fig. S4(d) is significantly different from the experimental results shown in Fig.3 in the main text, which also suggests that the zero-dispersion solitons (two-peak structures) observed in our study are dominated by the fifth-order dispersion. This is further verified by the collapsed snaking bifurcation in Fig. 4 in the main paper.

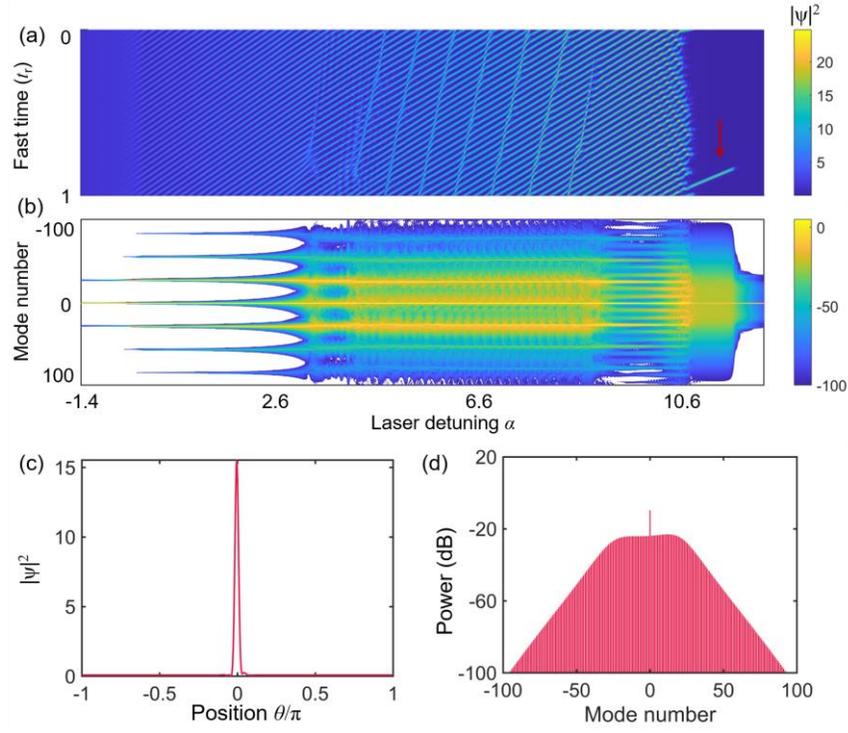

**Fig. S4** Numerical simulation of the comb dynamics with the impact of fourth-order dispersion. The simulation parameters are $D_2/2\pi = -390.4$ kHz, $D_3/2\pi = 19.9$ kHz, $D_4/2\pi = 9.5$ kHz, $D_5/2\pi = 0$ kHz, and $|F|^2 = 12$. Temporal (a) and spectral (b) evolution of the intracavity field while scanning the laser detuning from the blue- to the red-detuned side of the resonance. (c) and (d) show the temporal waveform and the corresponding spectrum at the detuning marked with an arrow in (a).

**Impact of fifth-order dispersion**

As shown in Fig. 4(a) in the main text, with the influence of the fifth-order dispersion, the existence range of multi-peak bright solitons (solid lines at different power level) is significantly broadened. Figure 4(a) also suggests that, with the influence of the 5$^{th}$-order dispersion, multi-peak soliton states with different temporal and spectral profiles can coexist under the same system parameters. Figure S5 shows numerical simulation results of three different soliton states with the same pump power ($|F|^2 = 15$) and laser detuning ($\alpha = 8.0$). These results are consistent with Fig. 4(a) and (d). Although beyond the scope of this work, we find that the inner peaks (marked with arrows) in Fig. S5 (c) and (e) are broader than the outer peaks, and look like multiple peaks closely bounded. Note that the numerical simulation in Fig. S5 is carried out with a fixed laser detuning, which corresponds to experimentally locking the laser detuning, similar to the work in Ref. [6]. Via scanning the laser detuning, apart from the two-peak soliton (Fig. 3 in the main text), other multi-peak soliton states are not experimentally observed. Figure S6 presents an example of numerically simulated intracavity dynamics (temporal (a) and spectral (b)) when slowly sweeping the laser frequency from the blue- to the red-detuned side of the resonance. The intracavity field experiences intense and random chaotic states and the spontaneous formation of bright soliton structures tends to be the state with the highest probability, which is the two-peak soliton in our case. Figure S6 (c) and (d) show the temporal profile and the corresponding optical spectrum of the final stable two-peak soliton at the detuning marked with an arrow in (a).

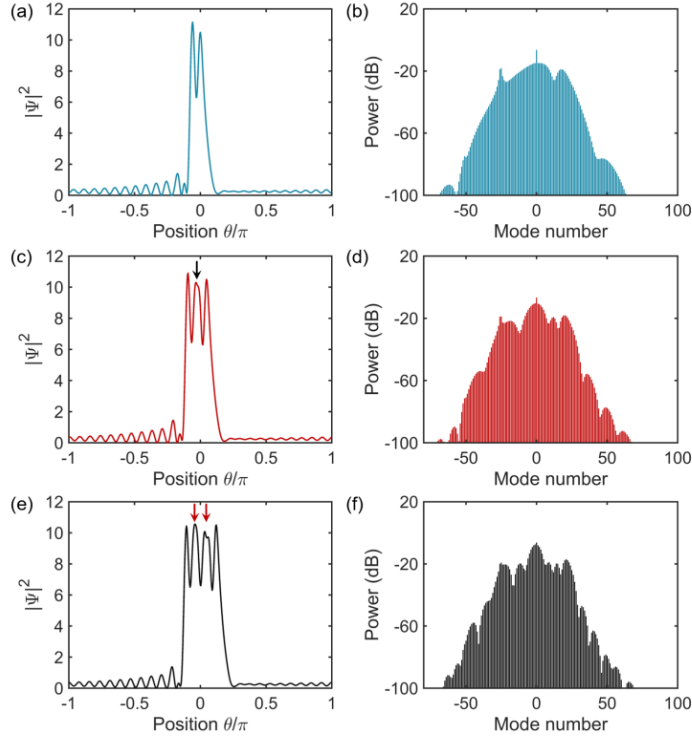

**Fig. S5** Different soliton states with different temporal (left column) and spectral profiles (right column) coexist under the same system parameters with fifth-order dispersion. The simulation parameters are $D_2/2\pi = -390.4$ kHz, $D_3/2\pi = 19.9$ kHz, $D_4/2\pi = 9.5$ kHz, $D_5/2\pi = 2.8$ kHz, $\alpha = 8.0$, and pump power $|F|^2 = 15$.

**Stability of a zero-GVD soliton**

The zero-GVD solitons driven by a CW source demonstrated in this work are stable and maintain their shape when changing the cavity-pump detuning, as long as the pump laser frequency is within the respective soliton step. Figure S7(a) shows the comb spectra measured at different cavity-pump detunings (the detuning is increasing from bottom to top within the panel). The spectral bandwidth increases with higher detuning, while the wavelength of the DW moves towards the longer wavelength from 1331 nm to 1343 nm. However, the position of the spectral dip on the left side of the pump laser remains almost fixed. Apart from these changes, the overall shape of the spectra remains constant, which suggests that the temporal waveform of the doublet soliton also maintains its temporal profile. We verify this finding by numerical simulations based on the LLE. Figure S7(b) shows simulated spectra at different cavity-pump detunings $\alpha$ (normalized to half of the full-width at half-maximum (FWHM) of the resonance), $\alpha = 6.1$, 7.1, and 8.5 for the bottom, middle, and top trace within the panel, respectively. In addition to the similar spectral envelopes, the simulated dip and DW in Fig. S7(b) also matches the behavior of the experimental results in Fig. S7(a). When the detuning increases, the DW shifts towards longer wavelengths while the spectral dip roughly remains in the same position. Figure 7(c) shows the evolution of the temporal waveform when the detuning changes from 7.9 to 9.7, further confirming that the temporal waveform stays stable. This observation agrees well with theoretical predictions [8].

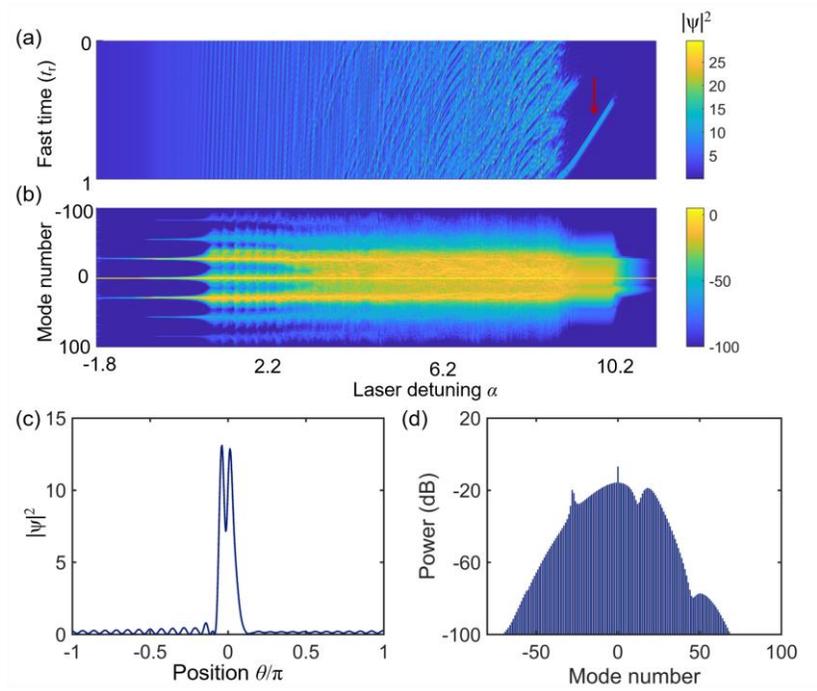

**Fig. S6** Numerical simulation of the comb dynamics including the fifth-order dispersion. The dispersion parameters are $D_2/2\pi = -390.4$ kHz, $D_3/2\pi = 19.9$ kHz, $D_4/2\pi = 9.5$ kHz, $D_5/2\pi = 2.8$ kHz). Temporal (a) and spectral (b) evolution of the intracavity field while scanning the laser detuning from the blue- to the red-detuned side of the resonance. (c) and (d) show the temporal waveform and the corresponding spectrum at the detuning marked with an arrow in (a).

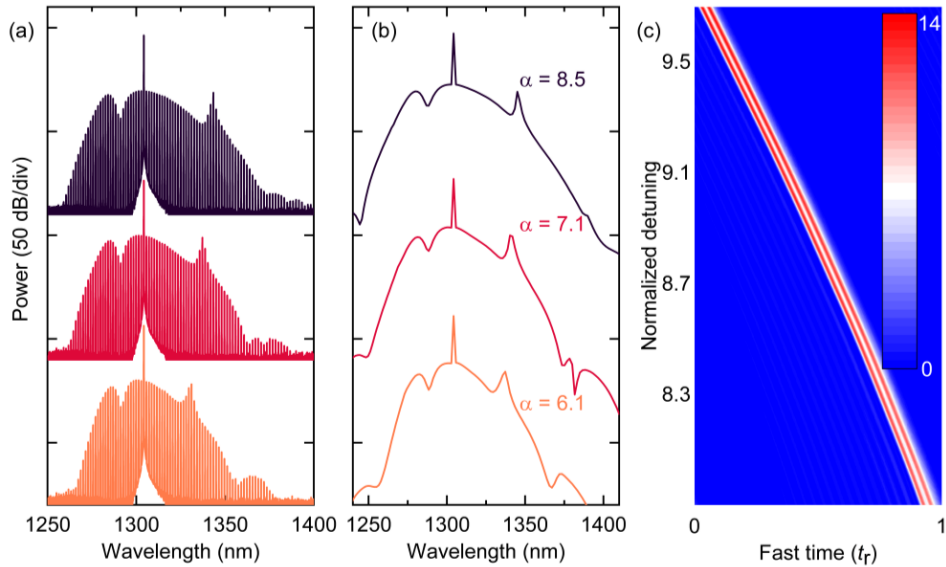

**Fig. S7** Experimental measurement and numerical simulations of zero-GVD soliton spectra at different cavity-pump detunings. (a) Experimental spectra with increasing detuning from bottom to top within the panel. (b) Simulated spectra at different normalized detunings α. (c) Temporal waveform evolution of the intracavity doublet soliton in the normalized detuning range from 7.9 to 9.7 half-linewidths of the optical mode.

**Optical spectrum of the two-peak soliton for autocorrelation measurements**

Figure S8 shows the optical spectrum that is recorded after sending the optical spectrum from Fig. 5(b) of the main text through the fiber Bragg grating (FBG) filter and subsequently amplifying it with a solid state amplifier (SOA) before sending it into the autocorrelator.

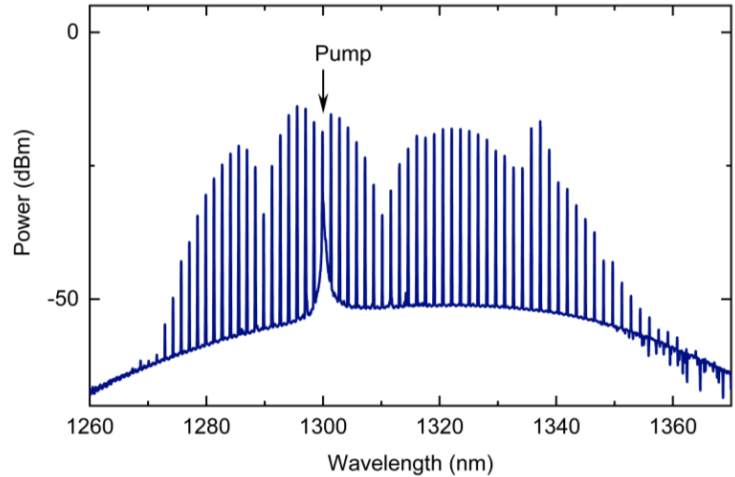

**Fig. S8** Optical spectrum after the FBG filter and after amplification by a solid state optical amplifier (SOA) prior to sending the signal into the autocorrelator. The arrow indicates the pump laser position.